\documentclass[11pt]{article}

\usepackage{amsmath}
\usepackage{amsfonts}
\usepackage{blindtext}
\usepackage{amsmath}
\usepackage{booktabs}
\usepackage{fancyhdr}
\usepackage{graphicx}
\usepackage{caption}
\usepackage{hyperref}
\usepackage{algpseudocode}
\usepackage{verbatim}
\usepackage{pdfpages}
\usepackage{listings}
\usepackage{mathtools}
\usepackage{tocloft}
\usepackage{amssymb}
\usepackage[ruled,linesnumbered]{algorithm2e}
\usepackage{animate}
\usepackage{afterpage}
\usepackage{pdfpages}
\usepackage{xcolor}
\usepackage{hyperref}
\begin{document}
\title{Source Separation of Multi-source Raw Music using a Residual Quantized Variational Autoencoder}

\author{
Leonardo Berti \\
Department of Computer Science, \\
Sapienza University of Rome
}
\maketitle
\begin{abstract}
\noindent I developed a neural audio codec model based on the residual quantized variational autoencoder architecture. 
I train the model on the Slakh2100 dataset, a standard dataset for musical source separation, composed of multi-track audio. The model can separate audio sources, achieving almost SoTA results with much less computing power.

\noindent The code is publicly available at \href{https://github.com/LeonardoBerti00/Source-Separation-of-Multi-source-Music-using-Residual-Quantizad-Variational-Autoencoder}{github.com/LeonardoBerti00/Source-Separation-of-Multi-source-Music-using-Residual-Quantizad-Variational-Autoencoder}
\end{abstract}

\section{Introduction}
In this work, I will focus on the source separation tasks of multi-source raw music.
I consider music in the raw audio domain in two different forms: as multiple individual sources $\{\mathbf{x}_1, ..., \mathbf{x}_n\}$ and as a mixture $\mathbf{y} = \sum_{i=1}^N \mathbf{x}_i$ where each source is represented as a continuous waveform $\mathbf{x} \in \mathbb{R}^T$, where $T$ is the product of the sample rate with the audio duration in seconds. 
Musical mixtures (stems) consist of sources that are contextually related and strongly dependent on each other. Therefore the joint distribution of the sources $p(\mathbf{x}_1, ..., \mathbf{x}_N)$ is not separable into the product of marginal distribution of each source $\{p_n(\mathbf{x}_n)_{n=1,...,N}\}$, this implies that going from the mixture to the individual sources is not an easy solvable problem. 

Generative State-of-The-Art (SoTA) audio models directly learn the distribution $p(\mathbf{y})$ of the mixtures, losing the information required for the separation task. 

In this work, I propose a residual vector quantized variational autoencoder (RQ-VAE) \cite{zeghidour2021soundstream} for the source separation task. I Train and evaluate the model on the the Slakh2100 dataset, a standard dataset for musical source separation, composed of 145 hours of multi-track audio.
The model achieves almost SoTA results with approximately one-hundredth of computing power. SoTA methods \cite{mariani2023multi, manilow2022improving} in source separation require a lot of computing power because they take advantage of a large number of inference steps, while RQ-VAE uses only one step.

% ------------------------------------------------------------------------------
\section{Related Work}

Audio source separation models can be divided into two main categories: discriminative and generative. Discriminative source separation models \cite{lluis2018end, luo2019conv, defossez2019music, takahashi2018mmdenselstm, choi2021lasaft, defossez2021hybrid} are deterministic parametric models that take the mixtures as input and output one or more sources, maximizing the likelihood of the conditional distribution of the individual sources given the mixture $p(\mathbf{x}_1, ..., \mathbf{x}_N | \mathbf{y})$.
These models are usually trained with a regression loss \cite{guso2022loss}, like L1 or L2, or signal-to-distortion ratio (SDR) \cite{vincent2006performance} on the reconstructed signal represented as waveform \cite{lluis2018end, luo2019conv, defossez2019music}, STFT \cite{takahashi2018mmdenselstm, choi2021lasaft} or both \cite{defossez2021hybrid}.   
Generative source separation models \cite{postolache2023latent, jayaram2021parallel, jayaram2020source, subakan2018generative}, instead, learn a model for each source, aiming at the marginal distribution of each source $\{p_n(\mathbf{x}_n)_{n=1,...,N}\}$. The mixture is only used during inference, where a likelihood function relates it to its component sources. Different types of priors have been investigated in the literature, such as GANs, normalizing flows, and autoregressive models.

% ------------------------------------------------------------------------------
\section{Method}\label{sec:latex}

\subsection{RQ-VAE}
In this section, I propose an RQ-VAE \cite{zeghidour2021soundstream, lee2022autoregressive} to perform source separation. The RQ-VAE compress the audio waveforms into hierarchical discrete representations in a lower-dimensional space. The residual vector quantizers consist of a hierarchy of Q vector quantizers, each composed of a codebook of N symbols. Hence after the quantization, the input audio samples are represented by a matrix $\mathbf{Z} \in \{1, ...., N\}^{T_c \times Q}$ with $T_c = T/f$, where f is the fold reduction. I configure the RQ-VAE to perform a 200-fold reduction in the sampling rate, producing embeddings at 110 Hz for input waveforms at 22 kHz, which results in a $\frac{22050 \times 16}{110 \times 12 \times 12} = 22$ bits compressing, where 110 is the encoded sample length, 12 are the bits used to represent each code (the codebooks are composed of $2^{12}=4096$) and 12 are the number of quantization layers used. The RQ-VAE takes in input the mixture and outputs the sources separated.
To try developing a general audio model, capable also of generating new music, I trained an RQTransformer \cite{lee2022autoregressive} in the latent space of the RQ-VAE but unfortunately, the quality of the music generated is unsatisfactory.
The details of the RQTransformer and other implementation details of the RQ-VAE are reported below.

\paragraph{Implementation details}The RQ-VAE consists of an encoder composed of four convolutional blocks followed by a bidirectional LSTM \cite{lstm} of two layers. Each convolutional block is composed of a strided convolution, followed by a residual block composed of three convolutional layers that maintain the dimensions unaltered and a residual connection. As striding factors, I use (5, 5, 5, 3). As kernel size I use (7, 7, 7, 5) Every time I use a striding convolution I double the channel size. 
The first convolutional block is preceded by a convolutional layer that augments the channel size to 8 with a kernel size of 3. All the convolutional layers used are 1D. As activation functions I used PReLU \cite{xu2015empirical} and as a regularization method I utilized batchnorm \cite{ioffe2015batch} in between every convolutional layer. The decoder is very similar to the encoder except for the fact that the strided convolutions are substituted with transposed convolutions. Between the encoder and the decoder I use skip connections as they have been shown to stabilize the training and augment the performance, this design choice led to an improvement of 1.7 in the SI-SDRI.
Inspired by \cite{zeghidour2021soundstream}, I enhanced the efficiency of the codebooks by initializing them with the centroids derived from executing the k-means algorithm on an encoded random training batch. This initialization strategy allows to start of the training with a high similarity of the codebook vectors with the audio encoding. Furthermore, I implemented a re-initialization method, when a codebook vector has not matched for several training batches I re-initialize it with a random input frame. More precisely, the exponential moving average of the assignments to each vector is tracked with a decay factor of 0.97 and the vectors are replaced if the assignment statistic falls below 2.
Different architectures and hyperparameters were tested. In particular, I tried to use dilated convolutions both in the residual blocks and strided convolutions, but it did not work, in fact, it diminished the performances. 
Another thing which I have experimented with is substituting the LSTM with self-attention blocks. 
Another relevant failed experiment was to try a NADE (Orderless Neural Autoregressive Density Estimator) approach to the training, which consists of estimating each source from both the mix and a random subset of the other sources, taking inspiration from \cite{manilow2022improving}, and using Gibbs sampling (up to 8 steps), this did not work probably for the small number of steps compared to the paper in which it was proposed, in fact the authors used 512 and starts to see improvement with respect of a more classical approach only after about 16 steps.
\paragraph{RQ-VAE Loss Function}
The loss function used to train an RQ-VAE to perform source separation is composed of three terms:
\begin{enumerate}
    \item Multi-scale spectral reconstruction term introduced in \cite{gritsenko2020spectral}, computed for each source:
    \begin{equation}
    \begin{split}
        \mathcal{L}^{spe}_R & =\frac{1}{6T} \sum_{source} \sum_{s \in 2^6, ..., 2^{11}} \sum_{t=1}^T ||\mathcal{S}^s_t(\mathbf{x}) - \mathcal{S}^s_t(\hat{\mathbf{x}})||_1 +\\
        & \alpha_s ||\log S_t^s(\mathbf{x}) - \log S_t^s(\hat{\mathbf{x}})||_2
     \end{split}
    \end{equation}
    where $\mathcal{S}^s_t(\mathbf{x})$ denotes the \textit{t}-th frame of a 64-bin mel-spectrogram computed with window length equal to $s$ and hop length equal to $s/4$. I set $\alpha_s$ = 1.
    \item Reconstruction term as mean squared error in the time domain, computed for each source:
    \begin{equation}
        \mathcal{L}^{rec}_R = \frac{1}{T} \sum_{source} \sum_{t=1}^T (\mathbf{x}_t - \hat{\mathbf{x}}_t)^2
    \end{equation}
    \item Commitment term to optimize the codebook symbols:
    \begin{equation}
    \begin{split}
        &\mathcal{L}^{comm}_R = \sum_{d=1}^D ||sg[z_e^d(\mathbf{x})] - z_q^d(\mathbf{x})||_2^2 + 
        \\ &\beta||z_e^d(\mathbf{x}) - sg[z_q^d(\mathbf{x})]||_2^2 + ||(z_e(\mathbf{x}) - sg[z_q^d(\mathbf{x})])||_2^2 
    \end{split}
    \end{equation}
    Where $z_q^d(\mathbf{x})$ is the quantized feature map at depth $d$, $z_e(\mathbf{x})$ is the encoder output before the quantization, $z_e^d(\mathbf{x})$ is the residuals at depth $d$ so $z_e^d(\mathbf{x}) = z_q^{d-1}(\mathbf{x}) - z_e^{d-1}(\mathbf{x})$, except for $d=1$, in that case is the encoder output.
\end{enumerate}
The overall autoencoder loss is a weighted sum of the different components to balance each term.

\paragraph{Hyperparameters search}
I have performed a hyperparameters search in this space:
batch size: $[8, \boldsymbol{16}, 32]$, learning rate: $[0.0001, \boldsymbol{0.001}]$, codebook length: $[512, \boldsymbol{4096}, 8192]$,
    number of quantization layers: $[8, \boldsymbol{12}]$, latent channel dimension: [128, $\boldsymbol{256}, 512], latent sample length: [294, \boldsymbol{441}]$. 
In bold are the best hyperparameters. As optimizer I used Adam. 
\subsection{RQTransformer}
I trained an RQTransformer to generate new music and get closer to a general audio model. The RQTransformer \cite{lee2022autoregressive} is trained in the latent space of the RQ-VAE.
After training the RQ-VAE, the latent codes are used as input to the RQTransformer to generate new music tracks. The RQTransformer is composed of a spatial transformer that takes in input the latent quantized representations and outputs a context vector that summarizes the information in previous positions, and a depth transformer that takes in input the context vector for the first position and then autoregressive predicts $D$ codes at position $t$. Both transformers are composed of stacked transformer encoder layers, composed of self-attention blocks and MLP. 
Generation is performed starting with a learnable start sequence and then generating the next codes latent in an autoregressive manner. For the configuration of the RQTransformer I followed the original paper \cite{lee2022autoregressive}. The number of depth transformer layers is equal to 4, the number of heads to 4, and the batch size is 16. 
I refer to the original implementation \cite{lee2022autoregressive} for more details. 
\paragraph{RQTransformer loss}
The RQTransformer is trained to minimize the negative log-likelihood:
\begin{equation}
    \mathcal{L}_{AR} = \mathbb{E}_S\mathbb{E}_{t,d}[-\log{p(S_{td}|S_{<t,d}, S_{t,<d})}]
\end{equation}
Where $S_{t,d}$ is the code at position $t$ of the $d$ quantization layers.

\section{Experiments}
\subsection{Experimental Setup}
To evaluate the RQ-VAE ability to separate sources, I use Slakh2100 \cite{manilow2019cutting}, a widely adopted data set for separation of musical sources. Slakh2100 consists of 2100 multi-track waveform music generated from MIDI files using high-quality virtual instruments. The data set is divided into 1,500, 375, and 225 songs for training, validation, and testing. Following \cite{manilow2022improving}, I focus on the four most common classes in the dataset: Bass, Drums, Guitar, and Piano. I selected Slakh2100 for its larger size with respect to other multi-source waveform datasets such as MusDB \cite{rafii2017musdb18}. 
I downsampled the audio tracks to 22kHz and used a context length of 4 seconds for training. The model was trained on an NVIDIA RTX 3090, for 60k steps, which correspond to 12 epochs.
\subsection{Results}
\subsubsection{Source Separation}
As evaluation metric I utilized the widely used Scale-Invariant Signal to Distortion Ratio improvement \cite{le2019sdr} SI-SDRi = SI-SDR$(\mathbf{x}_n, \hat{\mathbf{x}}_n)$, where $\mathbf{x}_n$ is the ground-truth source stem, $\hat{\mathbf{x}}_n$ is an estimate, and $\mathbf{y} = \sum_{i=1}^N \mathbf{x}_i$ is the mixture. 
I compare our method with 'Demucs + Gibbs (512 steps)' \cite{manilow2022improving} and MSDM \cite{mariani2023multi}, the state-of-the-art models in source separation on Slakh2100. These methods have in common the fact that they need hundreds of sampling steps to perform a single source separation step. Consequently, they need a lot of computing power and time, hundreds of times more than the RQ-VAE, which uses only one step.
I evaluated over the test set of Slakh2100, in the same way as in \cite{mariani2023multi, manilow2022improving}, using chunks of 4 seconds with at least two active sources and a two-second overlap.
    \begin{table}[!htb]\scriptsize
    \centering
    \caption{Quantitative results for source separation on the Slakh2100 test set. I use the SI-SDRi
as our evaluation metric (dB – higher is better). ‘All’ reports the average over the four stems (bass, drums, guitar and piano).}
    \label{tab:first_ex_table}
    \begin{tabular}{c|c}
    \toprule
    {\bf Model} & {\bf All} \\
    \midrule
    Demucs + Gibbs (512 steps) {\cite{manilow2022improving}} & 17.73 \\
    Weakly MSDM (correction) {\cite{mariani2023multi}} & 17.27 \\
    \midrule
    RQ-VAE & 11.49 \\
    \bottomrule
    \end{tabular}
    \end{table}
\subsubsection{Music Generation}
Unfortunately, the quality of the music generated is unsatisfactory. The negative log-likelihood reached in the test set is $5.342$, I cannot compare it with other models because the other models did not use the same loss.
I do not report other quantitative metrics, such as the Fréchet Audio Distance \cite{kilgour2019frechet} because in \cite{vinay2022evaluating} the authors empirically demonstrated that such metrics do not correlate with listener preferences. 
\section{Conclusion}
In this project, I developed a neural audio codec model, which relies on the residual quantized variational autoencoder architecture. The resulting model has the capability to separate audio sources, achieving almost SoTA results with much less computing power with respect of current SoTA methods. 
\bibliography{bibliography.bib}
\bibliographystyle{acm}
\appendix

\end{document}